\documentstyle[preprint,tighten,aps]{revtex}

\def\permille{$^\circ/_{\circ \circ} \;$}  

\def\lapprox{\mathrel{\mathop  {\hbox{\lower0.5ex\hbox{$\sim$}
\kern-0.8em\lower-0.7ex\hbox{$<$}}}}}  

\def\gapprox{\mathrel{\mathop  {\hbox{\lower0.5ex\hbox{$\sim$}
\kern-0.8em\lower-0.7ex\hbox{$>$}}}}}

\begin{document}

\draft

\preprint{\vbox{\noindent{}\hfill INFNFE-05-00}}

\title{Helioseismology and screening of nuclear reactions in the Sun}

\author{G.~Fiorentini$^{(1,2)}$, B.~Ricci$^{(1,2)}$ 
and F.L. Villante $^{(1,2)}$}

\address{
$^{(1)}$Dipartimento di Fisica dell'Universit\`a di Ferrara,I-44100
Ferrara,\\
$^{(2)}$Istituto Nazionale di Fisica Nucleare, Sezione di Ferrara, 
I-44100 Ferrara,}

\date{\today}

\maketitle

\begin{abstract}
We show that models for screening of nuclear reactions in the Sun
can be tested by means of helioseismology.
As well known, solar models using the weak screening
factors are in agreement with data. We find that the
solar model calculated with the anti screening factors
of Tsytovitch is not consistent with helioseismology,
both for the sound speed profile and for the depth of
the convective envelope. Moreover, the difference between the 
no-screening and weak screening model is
significant in comparison with helioseismic uncertainty.
In other words, the existence of screening can be proved
by means of helioseismology.
\end{abstract}

\section{Introduction}
\label{intro}

The solar neutrino problem is so important that any aspect of solar,
plasma and nuclear physics pertinent to
it has to be deeply investigated before
definitive conclusions can be drawn. In this respect, screening of the charges 
of the reacting nuclei due to free 
 charges in the solar plasma is of some interest.

 The study of screened nuclear reaction rates
 was started with the pioneering work of Salpeter 
\cite{salp}, who discussed both the extreme
 cases of "weak" and "strong" screening,
providing suitable expressions for the screening factors
\begin{equation}
\label{eqf}f_{ij}= \langle\sigma v\rangle _{ij,plasma} 
/ \langle \sigma v \rangle _{ij,bare} \, .
\end{equation}
The solar core is  not far from the weak screening case,
however it does not satisfy the usual conditions under which the
 weak screening approximation holds.
This is the reason why the problem has been investigated by several authors, 
see e.g. \cite{gdgc,koonin,mitler,dzitko,ricci,liolios}.

Gruzinov and Bahcall \cite{gb98} calculated the electron density in the
vicinity of fusing nuclei using the partial differential equation
for the density matrix that is derived in quantum statistical mechanics.
Their numerical result agrees, within small uncertainties, with Salpeter's
weak screening formula. Furthermore, Bahcall et al. \cite{b2000a}
recently provided several arguments that  demonstrate the validity of the
Salpeter formula near the solar center with insignificant errors.

The conclusions of Gruzinov and Bahcall \cite{gb98} are not unanimously
accepted. According to Shaviv and Shaviv \cite{s2}, the weak screening formula
does not hold in the Sun.  Some nuclear reactions  are enhanched
by the surrounding plasma whereas some others are suppressed.
According to  Tsytovitch and Bornatici \cite{tsy2000,bornatici} a
kinetic description of collective plasma effects results in
a decrease of the thermonuclear reaction rates in contrast 
to the Salpeter's enhancement.

We observe that solar models are built by using stellar evolutionary codes
which include specific expressions for the the nuclear reaction rates.
If one uses  different formulas for the screening factors $f_{ij}$
 one obtains different solar models.
On the other hand, helioseismology provides precise information
on the  sound speed profile and on the
properties of the convective envelope, see e.g. \cite{eliosnoi},
 which have to be reproduced by the correct solar model.
The main purpose of this paper is to test the screening models
by means of helioseismology. 
We build several solar models corresponding to different screening factors 
and  compare the results
with helioseismic data.

We also comment on the predicted neutrino fluxes.
Many of the attempts to modify the screening factors have been 
produced as an effort to avoid or mitigate the so called solar neutrino puzzle,
by reducing the predicted $^8B$ neutrino flux. We will show
that a reduction
 of the screening factors
does not  generally imply a reduction of the $^8B$ neutrino flux.

\section{Results of solar model calculations 
for different screening prescriptions}

\label{sec2}

By using  FRANEC  \cite{oscar}, a stellar evolutionary code
including diffusion of helium and heavy elements \cite{ciacio},
we constructed solar models based on four different assumptions:

\noindent  
i) The weak screening approximation (WES). The screening  factors $f_{ij}$ 
are given by:
\begin{equation}
\label{eqfweak}
\ln f_{ij}^{\rm{WES}} = Z_i Z_j e^{2} / (a_D\,kT)
\end{equation}
where $Z_i,Z_j$ are the charges of the interacting nuclei, 
$T$ is the temperature and $a_D$ is the Debye radius.
As clear from equation above,
 the screening factors are always larger than unity,
 i.e. the plasma provides enhancement of 
the thermonuclear reaction rates.

\noindent 
ii) The Mitler result \cite{mitler} (MIT), obtained with an analytical
method  which goes beyond the linearized approach and 
which correctly reproduces both the limits of weak and 
strong screening. Neglecting the small effects of a
radial dependence in the effective potential, see \cite{dzitko}, 
the enhancement factors are given now by:
\begin{equation}
\ln  f_{ij}^{\rm{MIT}}=-\frac{8}{5} (\pi e n_e a_D^5)^2 \,
[(\zeta_i+\zeta_j +1)^{5/3} - (\zeta_{i} +1)^{5/3} - 
(\zeta_{j}+1)^{5/3} ]/ (k T)
\end{equation}
where $\zeta_{i,j}=3 Z_{i,j}/4\pi n_e a_D^3$ and $n_e$ 
is the electron number density.

\noindent
iii) Neglect completely any screening effect (NOS),
 i.e. nuclear reactions 
occur with rates $\langle\sigma v\rangle_{\rm{bare}}$.
This case is considered in connection with  the suggestions 
that screening can be much smaller than Salpeter's estimate,
see e.g. \cite{s95}.

\noindent 
iv) The Tsytovich model (TSY) \cite{tsy2000,bornatici}, which
provides a decrease of all the thermonuclear reaction rates with 
respect to the case of bare nuclei.
Screening factors are taken from Table 1b of ref. \cite{bornatici}
and, for the $^7Be$ electron capture, from Table 2 of the same reference
(all factors are assumed constant along the solar profile).

In Table \ref{tabf} we report  the screening factors
 at the solar center for the various models.
One sees that  the weak screening approximation
 always yields the largest enhancement factors,
as physically clear due to the fact that electrons and ions are assumed
to be free and capable of following
the reacting  nuclei.
We also remind that  in this model the electron cloud is allowed to strongly
condense around the nuclei.
In the Mitler model, where electron 
density at the nuclear site is fixed at
$n_e$, the enhancement factor is smaller. By definition there is no
enhancement in the NOS model, whereas in TSY model there is a decrease
of the reaction rate, as already remarked.

The main features of the solar models we obtained are presented in
 Table \ref{tabmod}.
When moving from WES to solar models where nuclear reactions
are less favoured one observes the following effects:

i) The central temperature $T_{c}$ increases. 
In fact the hydrogen burning rate is fixed by the solar luminosity 
and a decrease of $f_{11}$ has to be compensated with a 
temperature increase;

ii) The isothermal sound speed, $u=P/\rho$ near the center increases.
This is due to the increase of temperature  whereas the ``mean molecular
weight''  remains approximately constant; 

iii) The properties of the convective envelope are affected. In particular
the border between the radiative and the convective region  moves
outwards and the photospheric helium abundance  decreases.

\section{Helioseismology and electron screening}

As well known several properties of the sun can be determined
accurately by helioseismic data, see e.g. \cite{valencia99,bpb2000}.  
The photospheric helium abundance $Y_{ph}$ and 
the depth of the convective zone $R_b$ are given by:
%
\begin{equation}
\label{eqyph}
Y_{ph}=0.249 (1\pm 1.4\%)
\end{equation}
%
\begin{equation}
\label{eqrb}
R_b/R_{\odot}=0.711 (1\pm 0.2\%) \, .
\end{equation}
%
The quoted errors are the so called  "statistical" or
"$1\sigma$"  errors
of \cite{eliosnoi,valencia99}. This error estimate was obtained 
by adding in quadrature each contribution to the uncertainty.
 Similar error estimates are given in \cite{ba95,ba97}.
A more conservative approach corresponds to 
add linearly all known individual uncertainties.
This gives the so called "conservative"  errors  
studied in \cite{eliosnoi}, which are about a factor
three larger than those in eqs. (\ref{eqyph},\ref{eqrb}).

Moreover, by inversion of helioseismic data one can determine
the sound speed profile in the solar interior.
This analysis can be performed either
in terms of the isothermal squared sound speed,
$u=P/\rho$, or in terms of the adiabatic squared sound speed
$c^2= \partial P/ \partial \rho|_{ad}=\gamma P/\rho$,
as the coefficient  $\gamma= \partial \log  P/ \partial \log  \rho|_{adiab}$
is extremely well determined by the equation of state of
the stellar plasma.
The typical "$1\sigma$" error on $u$ is about 
$1.3$ \permille in the intermediate solar region and increases
up to $7$ \permille near the solar center, see
\cite{eliosnoi}. A similar error estimate is obtained in \cite{bpb2000}.
The "conservative" error estimate of ref \cite{eliosnoi} is about
a factor three larger.

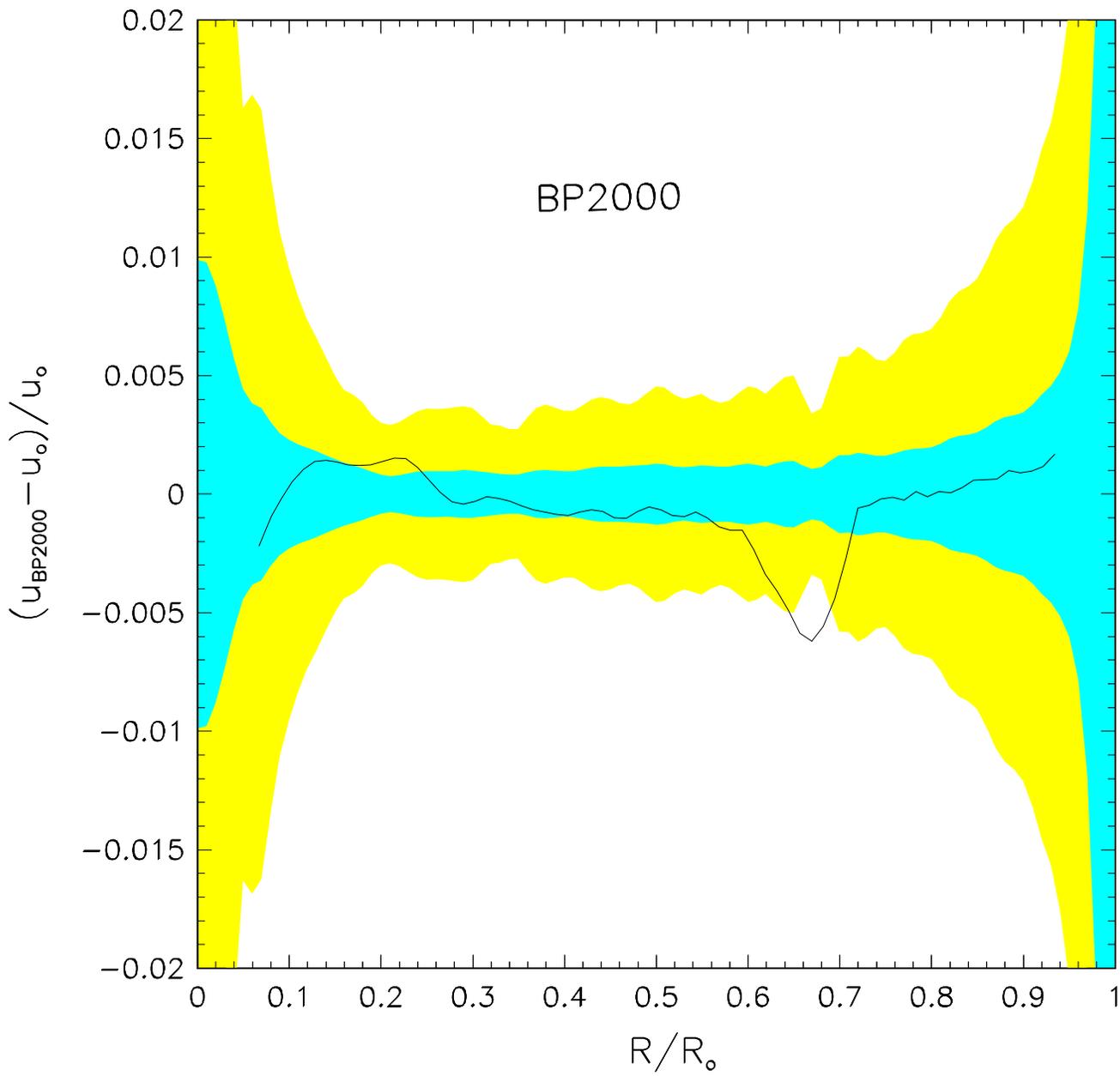
\begin{figure}
\caption[ubp98]{Comparison between the BP2000 model 
\cite{bp2000} and helioseismic data for $u=P/\rho$.
The ``statistical'' and ``conservative'' helioseismic uncertainties
\cite{eliosnoi} correspond to the dark and light areas
respectively.}
\label{figubp98}
\end{figure}

Recent Standard Solar Models calculated by using the weak screening
prescription are in agreement with   helioseismic constraints
on the properties of the convective envelope and on the sound speed
profile, see 
e.g. the BP98 model of ref \cite{bp98} and BP2000 model of
ref \cite{bp2000}.
As an example, we present in  Fig. \ref{figubp98}.
the comparison between the prediction of BP2000 and helioseismic data
for $u=p/\rho$.

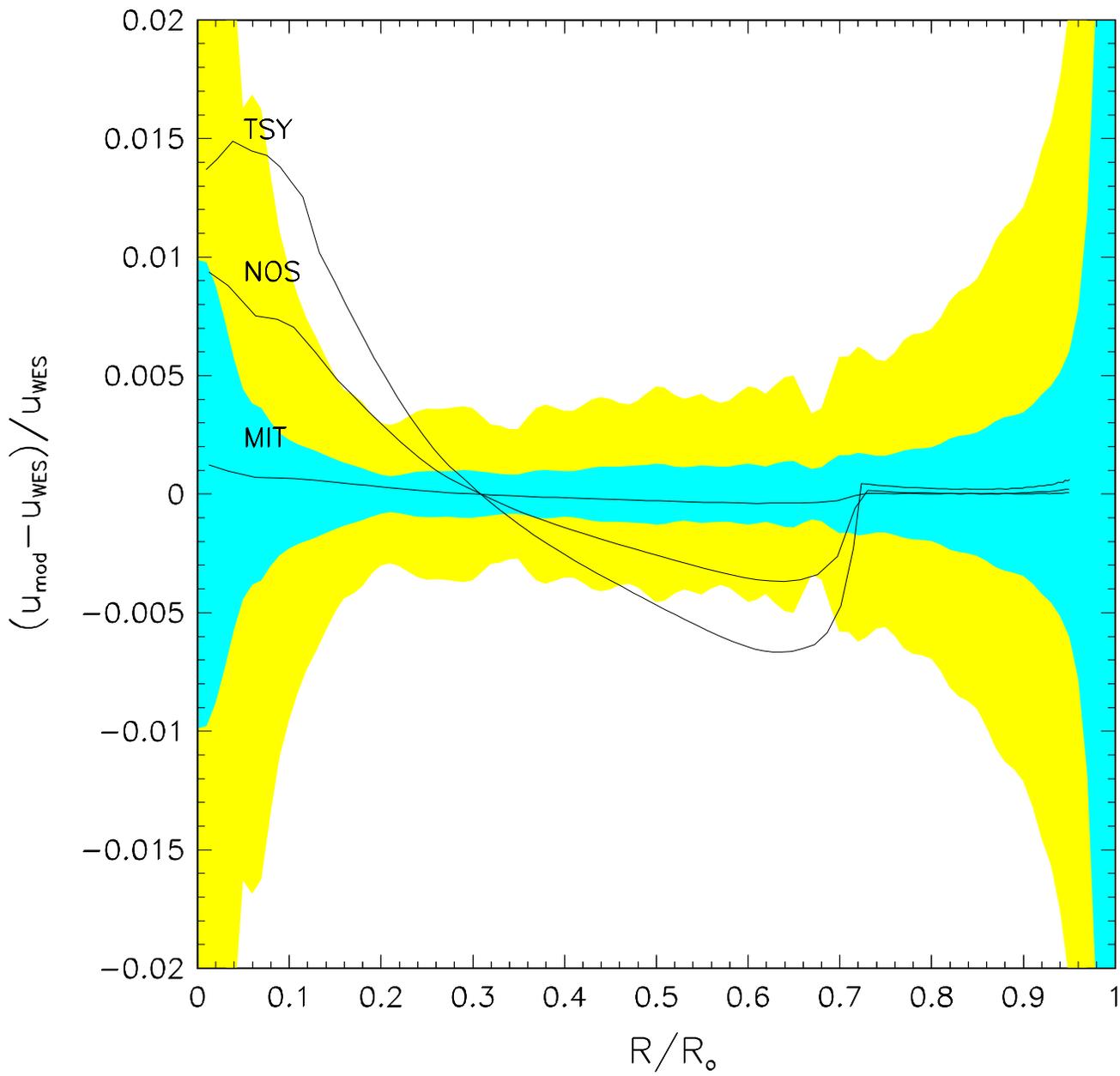
\begin{figure}
\caption[uprof]{Comparison of different screening models
with the WES model for  $u=P/\rho$, same notation 
as in Table I.
The ``statistical'' and ``conservative'' helioseismic uncertainties
\cite{eliosnoi} correspond to the dark and light areas.}
\label{figu}
\end{figure}

All this shows that the weak screening model is in agreeement with data
 and deviations from WES cannot be too large.
From the comparison of different models, see 
Table \ref{tabmod} and Fig \ref{figu}, one obtain the following results: 

i) The difference between the Tsytovitch model (TSY) and 
the weak screening model (WES) exceeds the ``conservative" 
uncertainty on $u$ in a significant portion of the solar profile.
We remark that also the depth of the convective envelope is
significantly altered. In other words the anti-screening predictions
of ref. \cite{tsy2000,bornatici}
can be excluded by means of helioseismology.

ii) Also the difference between the no-screening model (NOS) and WES 
is significant for both $u$ and $R_b$ in comparison with 
helioseismic uncertainty. In other words the existence of 
a screening effect can be proved by means of helioseismology. 

iii) The Mitler model of screening (MIT) cannot be distinguished
from the weak screening model within the present accuracy of helioseismology.

\section{Neutrino fluxes}
\label{neutrino}

A reduction of the screening factors does not automatically mitigate
the ``solar neutrino problem''. As an example
 the TSY model predicts a larger
$^8B$ flux, Chlorine and Gallium signals 
than the WES model, see last column of Tab. \ref{tabmod}.

As discussed extensively in ref \cite{ricci},
the behaviour of neutrino fluxes can be understood by considering
that a decrease of  the screening factors  has the 
following effects on the solar structure:

\noindent
i) The hydrogen burning rate is fixed by the solar luminosity 
and a decrease of $f_{11}$ has to be compensated with a 
temperature increase, being approximatively \cite{ricci}: 
\begin{equation}
T_{c}\propto f_{11}^{-1/8} \;\rm{;}
\end{equation}

\noindent
ii) The rate of the $^{3}He + ^{4}He$ reaction, 
which is responsible for the PP-II chain 
and for Beryllium neutrino production, is changed.
This results both from the increase in central temperature and 
from the variations in the screening factors $f_{34}$ and $f_{33}$, 
see \cite{ricci}. As a consequence, one expects a variation in the 
beryllium neutrino flux given by :
\begin{equation}
\label{ber}
\Phi_{Be}\propto \frac {f_{34}}{f_{33}^{1/2}} \cdot f_{11}^{-10/8}
\;\rm{,}
\end{equation}

\noindent
iii) The $^8B$ neutrino flux in addition depends on the
ratio of the proton to electron capture rates
on $^7Be$:
\begin{equation}
\Phi_{B}\propto \frac{f_{17}}{f_{e7}} 
\frac{f_{34}}{f_{33}^{1/2}} \cdot f_{11}^{-3} \;\rm{.}
\label{boro}
\end{equation}

These scaling laws account for the numerical results of 
Table \ref{tabmod}
and provide an explanation for the $^8B$ neutrino flux
increase of the TSY model.
As clear from eq.(\ref{boro}) effects on the proton
and electron capture  almost compensate 
($f_{17}^{WES}/f_{17}^{TSY}=2.213;\,
f_{e7}^{WES}/f_{e7}^{TSY}=2.166$),
and the increase  is essentially due to the $f_{11}^{-3}$
term, which corresponds to the temperature effect.

\section{Conclusions}
\label{conclusions}

We recall here the main points of our discussion:

\noindent
i)The anti-screening predictions of ref \cite{tsy2000,bornatici}
can be excluded by means of helioseismology, since both $u$ and $R_b$
are significantly altered.

\noindent
ii) We find that the  a  no-screening  solar model   
is not completely consistent with helioseismic data on $u$ and $R_b$,
in other words  the existence of 
a screening effect can be proved by means of helioseismology.


\begin{table}
\caption[aaaa]{
Screening factors in  solar center, for weak screening (WES) \cite{salp}, 
Mitler model (MIT) \cite{mitler}, no screening (NOS) and Tsytovitch model
 (TSY) \cite{tsy2000}.
}
\begin{tabular}{lr@{}lr@{}lr@{}lr@{}lr@{}lr@{}}
&\multicolumn{2}{c}{{}{}{}}
&\multicolumn{2}{c}{WES}
     &\multicolumn{2}{c}{MIT}
       &\multicolumn{2}{c}{NOS}
           &\multicolumn{2}{c}{TSY}\\
\tableline
$p+p$        &~~ &~~~                     &   1. & 049
   &        1. &045                &    &1 
   &        0. &949              \\
$^3He+^3He$       &~~ &~~~                       &  1. & 213
   &        1. &176                &   &1
   &        0. &814            \\
$^3He+^4He$       &~~ &~~~                       &  1. & 213
   &        1. &176                &   &1
   &        0. & 810            \\
$^7Be+p$       &~~ &~~~                       &   1. & 213
   &        1.  &171                &    &1 
   &        0.  &542            \\
\end{tabular}
\label{tabf}
\end{table}
\

\begin{table}
\caption[aaaa]{
Comparison among solar  models with different screening 
factors. We show the  fractional differences, (model -WES )/WES,
for the photospheric helium abundance ($Y_{ph}$), depth of the convective
envelope ($R_b$), central temperature ($T_c$), isothermal
sound speed squared at the solar center ($u_c$),
 neutrino fluxes ($\Phi_i$)
and predicted signals for Chlorine (Cl) and the Gallium (Ga) experiments.
\underline {All variations are in per cent}.}
\begin{tabular}{lr@{}lr@{}lr@{}lr@{}}
&\multicolumn{2}{c}{MIT}
  &\multicolumn{2}{c}{NOS}
     &\multicolumn{2}{l}{TSY} \\
\tableline
$Y_{ph}$                           &    -0.&076 
   &        -0.& 86               &   -1. & 4 \\
$R_b$                               &  + 0.& 037
   &        +0. &34              &   +0.& 59 \\
$T_c$                               &  + 0.&45
   &       +0. & 54               & +1.& 4 \\
$u_c$                               &  + 0.&10
   &       +1. & 0               & +1.& 4 \\
\tableline
$\Phi_{pp}$                                &  +0.& 033
   &       +0. &45               & -0. &35  \\
$\Phi_{^7Be}$                              &   -0. & 19
   &        -2.  &4          &   -5.  &9 \\
$\Phi_{^8B}$                             &   -2. & 7
   &        -12. &                  &  +11 &   \\
\tableline
Cl                                  &   -2. & 5
   &         -11. &                  & +9.&7 \\
Ga                                  & -0. &76
   &       -2. &9               & +2. &3 \\
\end{tabular}
\label{tabmod}
\end{table}

\end{document}